\documentstyle[12pt]{article}

\advance\textheight by \topskip
\newcommand{\sla}{\kern -5.4pt /}
\newcommand{\Dir}{\kern -6.4pt\Big{/}}
\newcommand{\Dirin}{\kern -10.4pt\Big{/}\kern 4.4pt}
\newcommand{\DDir}{\kern -7.6pt\Big{/}}
\newcommand{\DGir}{\kern -6.0pt\Big{/}}

\newcommand{\be}{\begin{equation}}
\newcommand{\ee}{\end{equation}}
\newcommand{\bea}{\begin{eqnarray}}
\newcommand{\eea}{\end{eqnarray}}
\newcommand{\beanon}{\begin{eqnarray*}}
\newcommand{\eeanon}{\end{eqnarray*}}
\newcommand{\ba}{\begin{array}}
\newcommand{\ea}{\end{array}}
\newcommand{\bd}{\begin{description}}
\newcommand{\ed}{\end{description}}
\newcommand{\bi}{\begin{itemize}}
\newcommand{\ei}{\end{itemize}}
\newcommand{\ben}{\begin{enumerate}}
\newcommand{\een}{\end{enumerate}}
\newcommand{\bc}{\begin{center}}
\newcommand{\ec}{\end{center}}

\newcommand{\vsk}{\vskip 10 pt\noindent}

\def\pl #1 #2 #3 {{\it Phys.~Lett.} {\bf#1} (#2) #3}
\def\np #1 #2 #3 {{\it Nucl.~Phys.} {\bf#1} (#2) #3}
\def\zp #1 #2 #3 {{\it Z.~Phys.} {\bf#1} (#2) #3}
\def\pr #1 #2 #3 {{\it Phys.~Rev.} {\bf#1} (#2) #3}
\def\prep #1 #2 #3 {{\it Phys.~Rep.} {\bf#1} (#2) #3}
\def\prl #1 #2 #3 {{\it Phys.~Rev.~Lett.} {\bf#1} (#2) #3}
\def\intj #1 #2 #3 {{\it Int. J. Mod. Phys.} {\bf#1} (#2) #3}
\def\mpl #1 #2 #3 {{\it Mod.~Phys.~Lett.} {\bf#1} (#2) #3}
\def\rmp #1 #2 #3 {{\it Rev. Mod. Phys.} {\bf#1} (#2) #3}
\def\cpc #1 #2 #3 {{\it Comp. Phys. Commun.} {\bf#1} (#2) #3}
\def\app #1 #2 #3 {{\it Acta Phys. Pol.} {\bf#1} (#2) #3}
\def\xx #1 #2 #3 {{\bf#1}, (#2) #3}

\begin{document}
\tolerance=100000
\thispagestyle{empty}
\setcounter{page}{0}

\begin{flushright}
{\large DFTT 59/97}\\
{\rm September 1997\hspace*{.5 truecm}}\\
\end{flushright}

\vspace*{\fill}

\bc
{\Large \bf \noindent
Reconstruction and approximation effects on W mass distributions at LEP2.
\footnote{ Work supported in part by Ministero
dell' Universit\`a e della Ricerca Scientifica.\\[2 mm]
e-mail: ballestrero@to.infn.it,chierici@to.infn.it}}\\[2.cm]

{\large  Alessandro Ballestrero$^{ab}$ and Roberto Chierici$^{ac}$}\\[.3 cm]
\ec
\vsk

\hskip 1.5cm $^a${\it I.N.F.N., Sezione di Torino, v. Giuria 1, 10125 Torino,Italy}

\hskip 1.5cm $^b${\it Dipartimento di Fisica Teorica, Universit\`a di Torino,
Italy}

\hskip 1.5cm $^c${\it Dipartimento di Fisica Sperimentale, Universit\`a di Torino,
 Italy.}

\vspace*{\fill}

\begin{abstract}
{\normalsize
\noindent
We analyze from a theoretical point of view the impact of various
approximations on W mass distributions. This is done both at parton level
and after a simulated W mass reconstruction using 
constrained fits. The results
may help to understand the origin of various shifts and the broadening of
the peak in direct reconstruction mass measurements.
 }
\end{abstract}
.  
\vspace*{\fill}
\newpage

\section{Introduction.}

One of the main purposes of LEP2 physics is the measurement of the W mass \cite
{yrmw}. 
The combined theoretical and experimental effort
will hopefully lead to its determination with an accuracy of the order of
50 MeV. After the threshold measurements at 161 GeV 
\cite{exp161},
in the runs at higher energies the so called direct reconstruction method will
be used. The first results at 172 GeV have already been reported by LEP 
collaborations \cite{othconf}\cite{delphiconf}. 
The above method is based on the determination
of the invariant mass of the two couples of fermions 
coming from the decay of the W's intermediate states. 
This is complicated by the initial state 
radiation (ISR), by the presence of the
invisible neutrinos in semileptonic channels and by the experimental difficulty
of reconstructing the invariant mass of two jets. 
For this reason kinematical constrained fits to the measured four-momenta
are used, in order to improve the resolution on the invariant mass
 \cite{othconf}\cite{delphiconf}. 
Another difficulty comes from the fact that  not only the three double resonant
diagrams (CC3) contribute to the four fermion final states, but also
 other diagrams which represent the so called irreducible background.
For instance the typical four quarks charged current process $u\bar d s \bar c$
(CC11) is composed of 11 diagrams and the semileptonic $e\bar \nu u\bar d$ 
(CC20) by 20. Moreover, as the light quark
flavour is practically indistinguishable, all four quark final states 
have to be accounted for and not only those coming from charged currents.
Many theoretical groups have  produced programs to compute four 
fermion processes (for a review of these generators, see \cite{yreg}). 
Some of these programs can compute all
possible processes and  a complete analysis of the various contributions to
invariant mass distributions has already been performed \cite{dis}.

In this letter we want to consider some of  the most relevant 
approximations and uncertainties which are  often unavoidable in the simulations
 and in the direct reconstruction method. In particular we want to study
their effect both at the parton level and after a simulated reconstruction
procedure in order to understand their relative importance on mass shifts
and distributions broadening. We do this with the help of WPHACT 
\cite{wphact}, one of the complete four fermion codes. 
To simulate the reconstruction, we  use a smearing of the theoretical
four momenta produced by WPHACT  with a procedure inspired by 
ref.~\cite{delphiconf} and then we  pass them to 
the constrained fit program PUFITC (for a description of the fitting technique
 see again ref.~\cite{delphiconf}).
The resulting distributions at generator (WPHACT) level and after the
simulated reconstruction are analyzed and fitted with a Breit Wigner (BW) to
determine a measure of the mass shifts and broadenings.

The plan of the paper is the following: in section 2 we give some details
on the distributions, their BW fits and estimated errors. In section 3 we
analyze the theoretical distributions at generator level. Section 4 is
dedicated to an explanation of smearing and constrained fit procedure.
The results for the  distributions and the parameters 
obtained in this way are discussed in section~5. Some conclusions are drawn
in section~6.

For the numerical part we have chosen as input parameters:
\[ m_W=80.356\, GeV, \quad m_Z=91.888\, GeV,  \quad 
s_{_W}^2 = 1 - {{m_W^2}\over {m_Z^2}}, \quad
g^2 = 4{\sqrt 2} G_\mu m_W^2. \]

The only cuts applied at parton level are those on the angle of each 
fermion with respect to the beam, which we have required to be greater than
10 degrees. Other cuts after momentum smearing are described in sect.~4.

\section {Distributions and their fit.}

 All distributions  were  produced using 
WPHACT with a binning of 100 MeV  and a relative accuracy per bin of 
a few per mill. To assess the differences
between mass distributions, we fitted them with a Breit-Wigner
corresponding to a running width W propagator squared:
\be\label{bw}
BW(m;m_W,\Gamma,N)=\frac{N}{(m^2-m_W^2)^2+(\Gamma m^2/m_W)^2},
\ee 
where $m_W,\, \Gamma$ and $N$  are the three parameters of our $\chi^2$ fit.

The difference between the values of the fitted BW masses and widths of
two curves can be considered as an indication of their relative
distortion.

All fits were  performed with PAW \cite{PAW} in the
interval 78 - 82.5 GeV. 
Such an interval was chosen in order not to fit the tails of the
distributions, where  phase space and matrix element effects are
stronger.

The $\chi^2/dof$'s obtained from the fit were generally  bad.
This is due to  the smallness of the errors on the distributions
and the fact that the Breit-Wigner curve is a very simple approximation
to the differential cross-sections $d\sigma/dm$.
This  is confirmed by  the sensitivity of
the parameters to the width
of the interval chosen. Varying it, for instance, to 78 - 82 GeV, 
produces a variation on the fitted mass  of some MeV.

We therefore conservatively associate an error to the fitted values of 
$m$ and $\Gamma$ which is rescaled with respect to the one given by PAW,
by a factor $\sqrt{\chi^2/dof}$. With such a procedure we obtain
a maximum error of 5 MeV for the mass. As far as the width is concerned,
one gets  a maximum error of  15 MeV for 
 parton level distributions  and 25 MeV for the others. 
These errors are however overestimated by about a factor 3 if they refer to the
definite interval used. From now on
it must be understood that the numerical results are affected by such
uncertainties.
 
With these provisos  we believe  that our fitted values and especially their 
differences can be considered good estimates of the shifts and broadenings.

\section {Parton level analysis of some approximations and constraints.}

We  study in this section the relevance on parton level computations
of some approximations and constraints which are sometimes used in simulations
or in the reconstruction of the W mass from experimental data. 
We   examine, for instance, the effect of neglecting ISR or that of computing
only the CC3 subset of diagrams. Such computations have  already been
performed many times in the literature
\cite{dis}\cite{many}\cite{yreg}\cite{yrww}.
However we do not   simply apply them to  total cross sections,
but to 
invariant mass  distributions, in order to understand how the
above approximations affect their maximum,  width  and eventually
 their shape.

\begin{figure}
\vspace{0.1cm}
\begin{center}
\unitlength 1cm
\begin{picture}(10.,5.)
\put(-1.5,-4.){\includegraphics{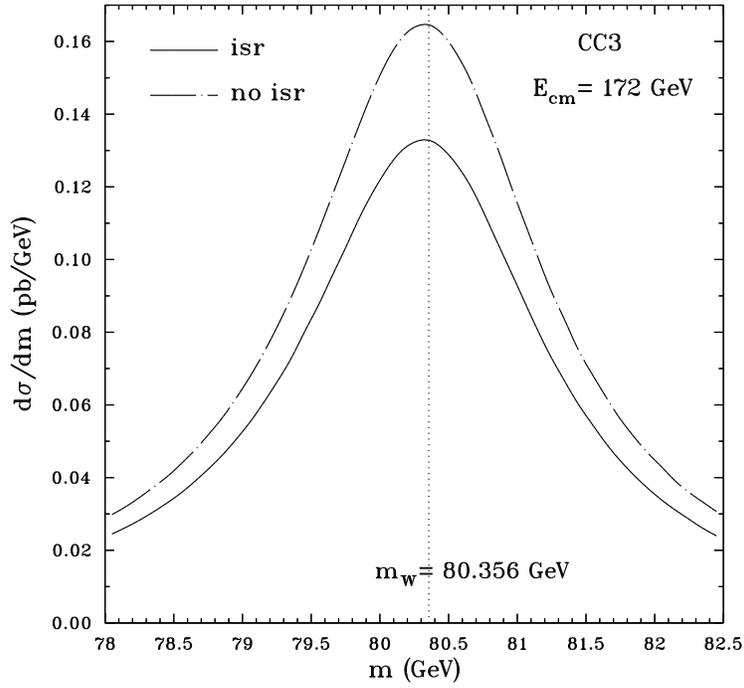}}
\end{picture}
\end{center}
\vspace{0.1cm}
\caption[]{ Invariant $u\bar d$ mass distribution as given by CC3 diagrams with
(continuous) and without (chaindot) ISR.  
}
\label{f1}
\end{figure}

Let us start  considering CC3  with and without ISR. 
An example of these distributions is reported in fig.~\ref{f1}.

The results of BW fits are reported in table~\ref{t1}. 

\begin{table}[hbt]\centering
\begin{tabular}{|c|c|c|c|c|c|c|}
\cline{2-7}
\multicolumn{1}{} {}&\multicolumn{2}{|c|}{172 GeV}& 
\multicolumn{2}{|c|}{184 GeV} &\multicolumn{2}{|c|}{200 GeV}\\
\cline{2-7} 
\multicolumn{1}{c|} {}&$m_W$     &  $\Gamma$   &$m_W$     
&  $\Gamma$   &$m_W$     &  $\Gamma$ \\
\hline
no ISR   &80.320 &2.062 &80.377  &2.087       &80.397 &2.091      \\
\hline
ISR      &80.296 &2.060 &80.364  &2.091       &80.389 &2.096       \\
\hline
\end {tabular}
\caption{ CC3 masses and widths (in MeV) with and without ISR for the average
distributions. }
\label{t1}
\end {table}

They refer to the distribution of the mean value
of the two reconstructed masses event by event. We will name it in the 
following average distribution, and it must not be confused with 
the one obtained taking the average of the two (e.g $m(e\bar\nu)$ and $m(u\bar d)$)
invariant mass distributions.
One notices a deviation of the fitted masses
from the input value $m_W=80.356$ GeV. At 172 GeV the values are lower and this
might be explained by the vicinity of the kinematical limit, while they
become larger at higher energies.  The shift due to ISR is of the order of 
25 MeV at
172 GeV and it decreases to 8 MeV at higher energies.  There is practically
no effect on the width. Coming back to fig~\ref{f1}, one may then say that
the major effect at the parton level of ISR on mass distributions is to 
lower them: this can also be proven by normalizing 
the two curves to the same value and superimposing them.

\begin{figure}
\vspace{1cm}
\begin{center}
\unitlength 1cm
\begin{picture}(10.,5.)
\put(-1.5,-4.){\includegraphics{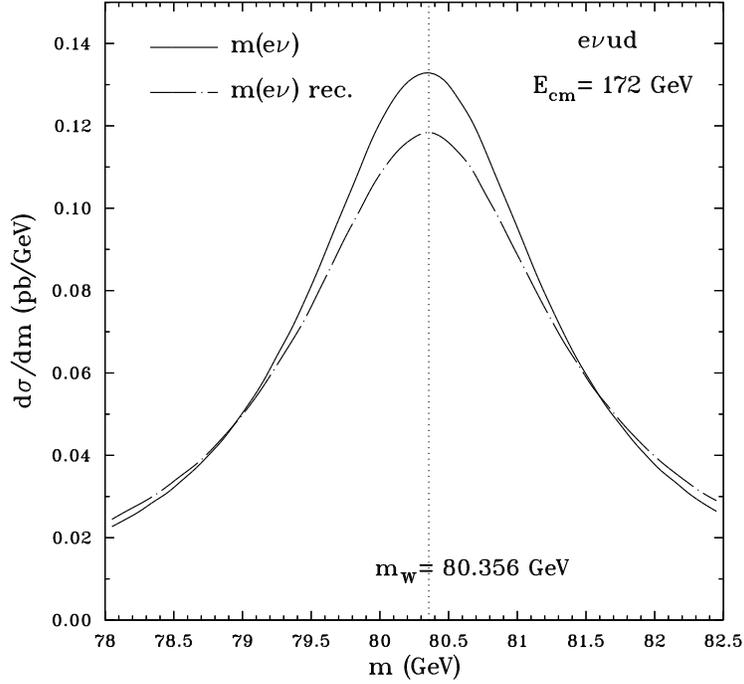}}
\end{picture}
\end{center}
\vspace{0.1cm}
\caption[]{ Invariant $e\bar\nu$ mass distribution for the complete 
$e\bar\nu u\bar d$
process. The continuous line corresponds to the true neutrino momentum, the
chaindot to the reconstructed one. 
}
\label{f2}
\end{figure}

There are  other ways in which ISR influences the mass measurements.
In the semileptonic final states, the neutrino
is invisible and its momentum can be only approximately reconstructed. The best
way to do it is to attribute all missing three momentum 
$\vec{p}_{mis}$ to the neutrino 
and take its energy to be equal to the modulus  $|\vec{p}_{mis}|$. 
This
method, which would of course be exact in case of no initial state 
radiation, leads to a distortion of the mass distribution. An example of this 
effect is shown in fig.~\ref{f2}. 
In table~\ref{t2} are reported the results obtained fitting with a 
BW $m(e\bar\nu)$ and average distributions.
$m$ and $\Gamma$ are the parameters resulting from fits 
with exact neutrino momenta. $\Delta m$ and $\Delta \Gamma$ are the differences
between the previous ones and those from fits to reconstructed neutrino
distributions. 

\begin{table}[hbt]\centering
\begin{tabular}{|c|c|c|c|c|c|c|c|c|c|c|c|c|}
\cline{2-13}
\multicolumn{1}{} {} & \multicolumn{4}{|c|}{172 GeV}& 
\multicolumn{4}{|c|}{184 GeV} &\multicolumn{4}{|c|}{200 GeV}\\
\cline{2-13} 
\multicolumn{1}{c|} {} & m & $\Gamma$& $\Delta m$ & $\Delta\Gamma$ 
& m & $\Gamma$& $\Delta m$ & $\Delta\Gamma$ 
& m & $\Gamma$& $\Delta m$ & $\Delta\Gamma$ \\
\hline
$e\bar\nu$ & 80376 & 2096 & 19 & 255 
       & 80408 & 2100 & 19 & 326 
       & 80417 & 2101 & 18 & 364 \\
\hline
Av & 80340 & 2074 & 24 & 180 
   & 80405 & 2104 & 25 & 269
   & 80426 & 2108 & 24 & 322 \\
\hline
\end {tabular}
\caption{Masses, widths and their differences (in MeV) between results from
 reconstructed and exact neutrino momenta.
The data refer to CC20 $e\bar\nu$ and average invariant mass distributions.
}
\label{t2}
\end {table}

The results of the table show clearly that the shift for this reconstructed
momentum effect is of the order of 20 MeV, not dependent on the energy.
The width is enlarged of about 10\%.

In practice these reconstructed mass distributions are never used: the 
reconstructed neutrino momentum is  normally an input of a fit procedure 
in which the conservation of the energy and the equality of the leptonic
and hadronic invariant masses are used as constraints.
It is well known that in reality the two invariant masses are not equal 
event by event.  If we compute the distribution corresponding to  the 
difference between $e\bar\nu$ and $u\bar d$ invariant  masses in $e\bar\nu u
\bar d$ final state at 172 GeV, one finds \cite{crac} 
that its maximum at zero is about 1 pb, and its width at half height is about
4 GeV. 
The width of the two BW's, the contribution of non resonant diagrams and
kinematic effects produce such difference. 
Nevertheless, requiring equal masses in the fit represents an approximation 
which is useful to improve considerably the mass resolution~\cite{yrmw}.

\begin{figure}
\vspace{0.1cm}
\begin{center}
\unitlength 1cm
\begin{picture}(10.,5.)
\put(-1.5,-4.){\includegraphics{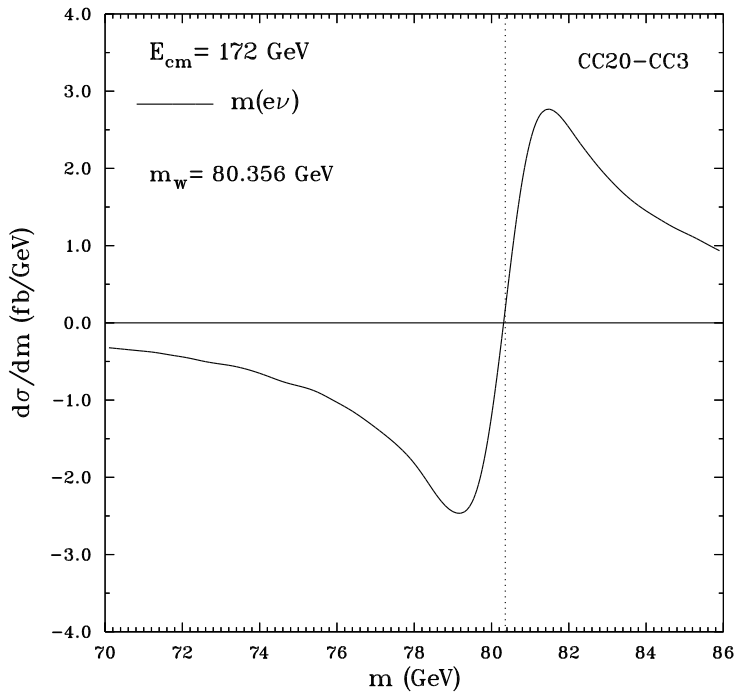}}
\end{picture}
\end{center}
\vspace{0.1cm}
\caption[]{ Difference between the invariant $e\bar\nu$ mass distribution 
produced by CC20 and CC3 diagrams.
}
\label{f3}
\end{figure}

We want now to understand how using complete calculations instead of double
resonant diagrams only, reflects on  fitted
masses and widths. To this end we have performed the  fit on 
CC3 and CC20 distributions and we report the results in table~\ref{f3}.

\begin{table}[hbt]\centering
\begin{tabular}{|c|c|c|c|c|c|c|c|c|c|c|c|c|}
\cline{2-13}
\multicolumn{1}{} {} & \multicolumn{4}{|c|}{172 GeV}& 
\multicolumn{4}{|c|}{184 GeV} &\multicolumn{4}{|c|}{200 GeV}\\
\cline{2-13} 
\multicolumn{1}{c|} {} & m & $\Gamma$& $\Delta m$ & $\Delta\Gamma$ 
& m & $\Gamma$& $\Delta m$ & $\Delta\Gamma$ 
& m & $\Gamma$& $\Delta m$ & $\Delta\Gamma$ \\
\hline
$u\bar d$ & 80332 & 2089 & -4 & -2 
   & 80366 & 2096 & -3 & 1 
   & 80379 & 2098 & 0  & 0 \\
\hline
$e\bar\nu$ & 80331 & 2090 & 44 & 6  
       & 80366 & 2096 & 42 & 4 
       & 80380 & 2097 & 37 & 3 \\
\hline
Av & 80296 & 2060 & 44 & 14 
   & 80364 & 2091 & 41 & 13 
   & 80389 & 2096 & 37 & 12 \\
\hline
\end {tabular}
\caption{CC3 masses, widths and  differences  between  CC20 and CC3
values (in MeV).
The data refer  $e\bar\nu$, $u\bar d$  and average invariant mass distributions.
}
\label{t3}
\end {table}

We do not show a similar table to compare CC11 
versus CC3,  simply because the fitted values  of masses and widths given by 
CC11 and CC3 are
always within their estimated errors.
 The same applies to CC10 ($\mu\bar\nu u \bar d$).
One immediately notices from table~\ref{t3} that also $\Delta m$'s and 
$\Delta \Gamma$'s for quark distributions are irrelevant.
The sizeable shift is on $\Delta m$'s for $e\bar\nu$ distribution. It
is of the order of 40 MeV and it is amazing that it practically
reflects  entirely on the average distribution.  
From fig.~\ref{f3} one realizes that the reason of the shift between
CC20 and CC3 is completely due to interference effects. The interference
between the double resonant diagrams and the diagrams which are obtained
from the CC10 ones with the exchange of  incoming $e^+$ with outgoing $e^-$,
has only one $W^-$ propagator. These contributions  change sign when 
$m(e\bar\nu)$ passes through $m_W$. For such a reason they depress
 masses lower than $m_W$
and enhance the higher ones, thus leading to the shift. This implicitly
makes us understand why CC20 $m(u\bar d)$ and CC10 or CC11, which  are not 
affected by 
such contributions, do not have a sizeable shift with respect to CC3.

\section{Kinematic reconstruction of the event: its effect on the 
mass distribution.}\label{kin}

To investigate the effects of a kinematic
reconstruction of the W resonance, the four-momenta generated with
WPHACT were first smeared to reproduce detector inefficiencies 
and hadronization effects and
then used in a constrained fit to simulate an experimental
determination of $m_W$.
The smearing applied to  three-momentum of the $i$-th quark 
was inspired by \cite{delphiconf} and can be written as:
\begin{equation}
\vec{p^i}_s = e^a\vec{p^i}_g+b\vec{p^i}_{perp,1}+b\vec{p^i}_{perp,2}
\end{equation}
where the subscripts $s$ and $g$ indicate respectively the smeared 
momenta and the
generated ones; $\vec{p^i}_{perp,1}$ and $\vec{p^i}_{perp,2}$ are two
versors orthogonal to $\vec{p^i}_g$ with random orientations. 
The variables $a$ and $b$ are random factors distributed according to
gaussians with central values $a_0$, $b_0$ and variances $\sigma^2_a$, 
$\sigma^2_b$. These parameters characterize the smearing and have
the following dependence on the polar angle of the parton:

\begin{center}
\begin{tabular}{ll}
$a_0$ = $-0.15-0.4\cos^6{\theta}$ & $\sigma_a$ = $-a_0$ \\
$b_0$ = 0 & $\sigma_b$ = 1 \\
\end{tabular} 
\end{center}
The electron and muon three-momenta were rescaled only in the
longitudinal direction
by smaller gaussian factors (with $\sigma$=0.07 for $e$ and $\sigma$=0.03 for
$\mu$), which were worsened in the forward regions, defined as 
$\theta <$40$^{\circ}$, $\theta >$140$^{\circ}$, where detection can
be less efficient ($\sigma$=.1 for $e$ and $\sigma$=.05 for $\mu$). 
The parton energies were then rescaled according to the three momentum
smearing. 

In order to consider only those events which allow acceptable
reconstruction, some selection cuts were applied after the smearing: 
to impose good jet-jet and lepton-jet separation,  
the minimum angle between two quarks was required to be greater 
than 5$^{\circ}$ and the minimum angle between the charged lepton 
and the jets to be greater than 10$^{\circ}$. The minimum invariant
mass of jets coming from the same W was 30 GeV.
In order then to guarantee the detection of the objects, only 
partons with polar angle in the region
10$^{\circ}<\theta <$170$^{\circ}$ were considered.   

The kinematic constrained fit was then performed on the smeared
four-momenta with PUFITC, applying four-momentum 
conservation and asking for two equal masses in the event. The number 
of constraints is
therefore five for fully hadronic events and two in the mixed
hadronic-leptonic ones: in the latter case, in fact, three of the
constraints must be used to determine the neutrino direction and momentum.
The input errors on jets and leptons for the fit were parametrized in
the same way as those used in the smearing procedure. 

 The effect of requiring equal masses
has already been mentioned before: for our purposes, the main 
result is that the observable to be compared to the fitted 
mass becomes the average W mass in the event.
One must notice that the procedure here described contains other 
approximations: detector effects leading to non gaussian errors on
jets and leptons are neglected, hard gluon radiation in the
final state leading to multi-jet structures and the
problem of the correct pairing of the jets in the fully hadronic 
channel are also ignored. 
Nevertheless the results we obtain give, in our opinion, 
a good indication of the effect of a kinematic fit on the mass
distributions. 

\begin{figure}
\vspace{1cm}
\begin{center}
\unitlength 1cm
\begin{picture}(10.,5.)
\put(-1.5,-4.){\includegraphics{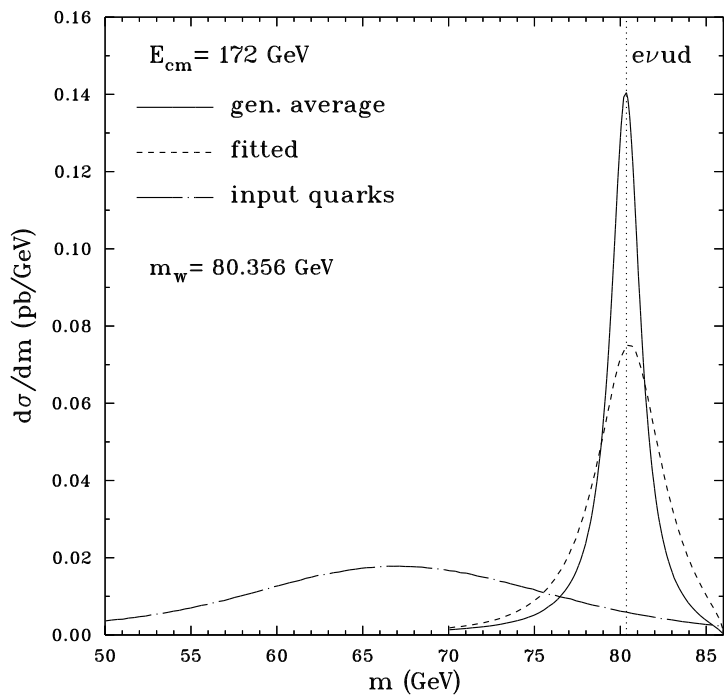}}
\end{picture}
\end{center}
\vspace{0.1cm}
\caption[]{ Invariant  mass distributions for the complete 
$e\bar \nu u \bar d$ process. 
The continuous line corresponds to the average 
(m($e\bar \nu$)+m($u\bar d$))/2 mass as generated
by WPHACT. The chaindot one corresponds to m($u\bar d$) after smearing. The
dashed line represents the distribution after the kinematical fit done with 
PUFITC.}
\label{f4}
\end{figure}

The three curves in fig.~\ref{f4} show the differential 
CC20 cross section 
distributions of the average mass in the event at 
generator level, after the application of the smearing and after the full
reconstruction of the event. The centre-of-mass energy chosen is 172 GeV.
It has to be remarked that the $d\sigma/dm$ distribution coming from the
constrained fit is a convolution of the generated differential
cross-section with the resolution function which, in our case, depends
only on parton level smearing. The resolution function is in 
general a complicated curve whose shape, in principle, depends on 
the mass itself and on the closeness to the kinematic limit. 
A test of this will be given in the next section. 

\begin{figure}
\vspace{0.1cm}
\begin{center}
\unitlength 1cm
\begin{picture}(10.,5.)
\put(-1.5,-4.){\includegraphics{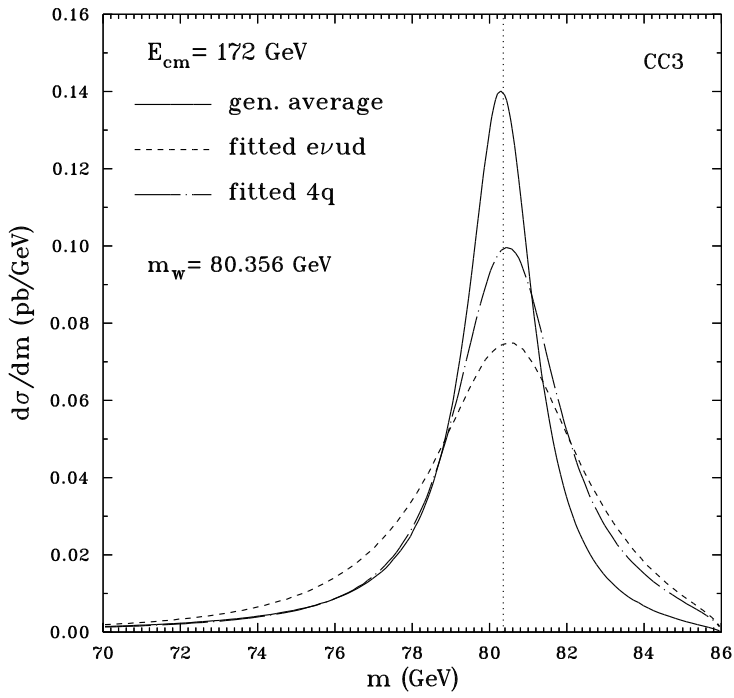}}
\end{picture}
\end{center}
\vspace{0.1cm}
\caption[]{Invariant  mass distributions for CC3 contribution.
The continuous line corresponds to the average 
(m($e\bar \nu$)+m($u\bar d$))/2 mass as generated
by WPHACT.  The dashed line represents the result of the fit with PUFITC
for a final $e\bar \nu u \bar d$ state. 
The chaindot one is the result of the same fit 
for a four quarks finale state.
}
\label{f5}
\end{figure}

Figure \ref{f5} presents CC3 $d\sigma/d\bar{m}$ distribution
in a different mass range, and the superimposed curves refer to the
reconstructed mass distributions in the $q\bar{q}q\bar{q}$ and
$q\bar{q}e\nu_e$ channels. One immediately notices the broadening of the 
two reconstructed curves and the fact that the one for the semileptonic 
curve is bigger. This difference is basically due to the loss of 
information caused by the missing neutrino. In the next
section we will quantify this difference in terms of shifts in mass
and width, trying to separate different contributions to
the distortions.

\section {Distortions in  distributions after constrained fit and 
their 
$\sqrt{s}$ 
dependence.}

In this section we study quantitatively the effects described above
after the smearing and the reconstruction with the constrained fit.

The reconstruction procedure is not much sensitive to
CC20 interference effects. Therefore no marked change, 
with respect to parton level results, is observed
in the CC20-CC3 shifts of the mass and of the width of the distributions. Their
value, determined by the difference of the parameters of Breit-Wigner
fits to the distributions are reported 
in table~\ref{tabcc3cc20}, for different values of the
centre-of-mass energy. 
The shifts, about 1.5 times bigger than  the
ones at generator level of sect.~3, do not depend strongly on ISR 
and can be considered almost constant in $\sqrt{s}$.

\begin{table}[hbt]\centering
\begin{tabular}{|c|c|c|c|c|c|c|}
\cline{2-7}
\multicolumn{1}{} {} & \multicolumn{2}{|c|}{172 GeV}& 
\multicolumn{2}{|c|}{184 GeV} &\multicolumn{2}{|c|}{200 GeV}\\
\cline{2-7} 
\multicolumn{1}{c|} {} & $\Delta m$ &  $\Delta \Gamma$   &
                         $\Delta m$ &  $\Delta \Gamma$   &
                         $\Delta m$ &  $\Delta \Gamma$ \\
\hline
no ISR & 63 & 19 & 68 & 35 & 63 & 34  \\
\hline
ISR    & 60 & -2 & 68 & 5 & 56 & -34   \\
\hline
\end {tabular}
\caption{Mass and width difference (in MeV) of distributions corresponding to
CC20 and CC3 $qqe\nu$ diagrams ($\Delta m=m_{CC20}-m_{CC3}$), 
for three values of 
\protect{$\sqrt{s}$}, 
with and without ISR.}
\label{tabcc3cc20}
\end {table}\

Table~\ref{tabfit} summarizes  the effects of
the kinematic reconstruction on the mass distribution in terms of mass and 
width shifts.  
From the numbers in the table it turns out that the reconstructed distributions 
 are broader and shifted towards higher masses.    
There are several causes for this behaviour, as we have discussed in
the previous section: missing energy coming from ISR, missing
energy coming from the neutrino in the semileptonic channel and
intrinsic resolution effects. The latter seem to play a
less important role in our framework with respect to the first two,
as can be seen from table~\ref{tabfit} by comparing
the shifts in the fully hadronic channel, no ISR,   with
the others. 
The basic effect of pure detector smearing is to broaden the mass 
distributions, as evident from the $\Delta \Gamma$'s, introducing relatively
small biases in the mass. 
The main
contribution to the shift in mass comes from the fact that in the constrained
fit the conservation of the energy tends to share the missing ISR energy
between the four-momenta in play, increasing the masses in the event.  
Another interesting information from the table is obtained by
comparing corresponding numbers at different centre-of-mass energies:
both shift in mass and broadening of the mass distributions increase.
This is mainly due to the combined effect of two causes:
 the first  is  the increase of the missing ISR energy
and the second is that the resolution function which convolves the
theoretical $d\sigma/dm$ distribution is broader because the 
kinematic limit moves away from the resonance and hence the 
reconstruction error on the mass becomes larger.

\begin{table}[hbt]\centering
\begin{tabular}{|c|c|c|c|c|c|c|c|}
\cline{3-8}
\multicolumn{1}{} {} & \multicolumn{1}{} {} &
\multicolumn{2}{|c|}{172 GeV}& 
\multicolumn{2}{|c|}{184 GeV} &\multicolumn{2}{|c|}{200 GeV}\\
\cline{3-8} 
\multicolumn{1}{c} {} & 
\multicolumn{1}{c|} {} & $\Delta m$ &  $\Delta \Gamma$   &
                         $\Delta m$ &  $\Delta \Gamma$   &
                         $\Delta m$ &  $\Delta \Gamma$ \\
\hline
& no ISR & 51 & 2083 & 129 & 3031 & 197 & 3862  \\
\raisebox{1.5ex}[0cm][0cm]{$e\bar\nu u\bar d$} 
& ISR    & 324 & 2427 & 443 & 3408 & 523 & 4234   \\
\hline
& no ISR & 5 & 757 & 38 & 1268 & 68 & 1746  \\
\raisebox{1.5ex}[0cm][0cm]{4q} 
& ISR    & 248 & 1072 & 305 & 1581 & 353 & 2069   \\
\hline

\end {tabular}
\caption{Mass and width difference (in MeV) between reconstructed mass and 
average 
generated one ($\Delta m=m_{rec}-m_{gen}$), 
for three values of 
\protect{$\sqrt{s}$}.
 The numbers
have been determined for $qqqq$ and $qqe\nu$ channels, with and without
ISR.}
\label{tabfit}
\end {table}

\begin{figure}
\vspace{1cm}
\begin{center}
\unitlength 1cm
\begin{picture}(10.,5.)
\put(-1.5,-4.){\includegraphics{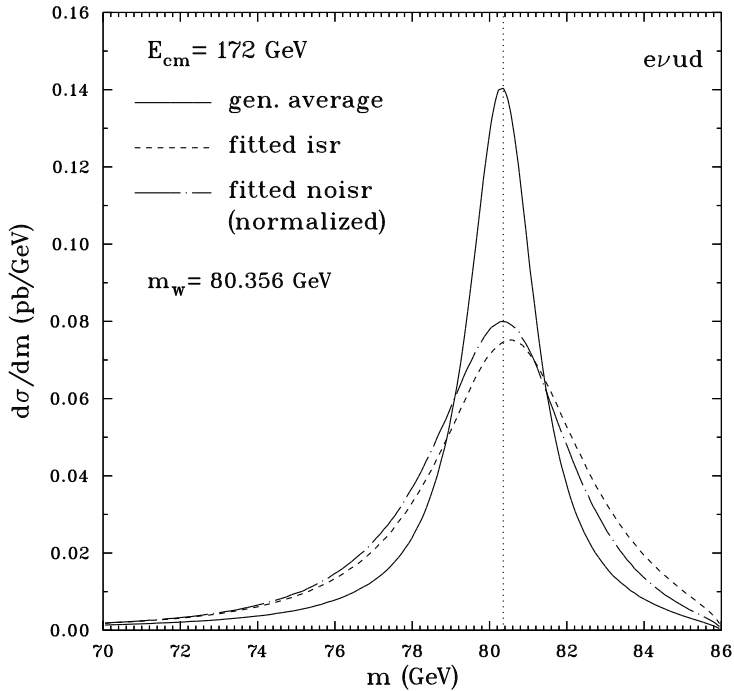}}
\end{picture}
\end{center}
\vspace{0.1cm}
\caption[]{Invariant  mass distributions for the complete $e\bar \nu u \bar d$ 
process.
The continuous line corresponds to the average 
(m($e\bar \nu$)+m($u\bar d$))/2 mass as generated
by WPHACT. The dashed line represents the result of the fit with PUFITC. 
The chaindot is the result of the same fit for the process with no ISR,
normalized to the ISR one.
}
\label{f6}
\end{figure}

We tried to further quantify the different contributions to the 
mass distortions: 
figure~\ref{f6} shows a comparison between the CC20 
average mass distribution produced by WPHACT (continuous line) with 
the reconstructed ones 
with (dashed) and without (chaindot) the presence of ISR.
The reconstructed mass distribution without ISR has been normalized to
the one with ISR. The comparison is interesting because it separates
the effect of ISR on  neutrino momentum reconstruction and on energy 
conservation in the fit from the distortions
coming from hadronization and detector smearing.
In case of
presence of ISR, as already discussed, its missing energy 
is associated to the
detected partons and therefore the invariant mass is artificially
increased with the result of a shift in mass. 

\begin{figure}
\vspace{0.1cm}
\begin{center}
\unitlength 1cm
\begin{picture}(10.,5.)
\put(-1.5,-4.){\includegraphics{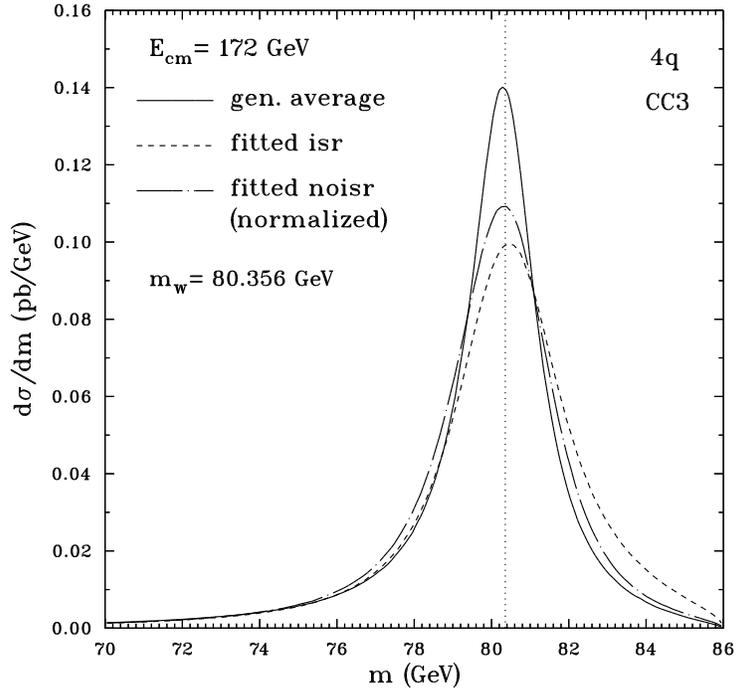}}
\end{picture}
\end{center}
\vspace{0.1cm}
\caption[]{ Invariant  mass distributions for CC3 contribution in a four quark
final state. The continuous line corresponds to the average 
mass as produced by WPHACT. The dashed line represents the result of the fit 
with PUFITC. The chaindot is the result of the same fit for the process with 
no ISR, normalized to the ISR one.
}
\label{f7}
\end{figure}

The missing neutrino reconstruction problem is absent in fig.~\ref{f7}, 
where ISR effect is shown in the reconstruction
of the fully hadronic channel. CC3 $d\sigma/d\bar{m}$ true and
reconstructed distributions are compared following the procedure of
fig.~\ref{f6}: in this case the difference between the dashed and
chaindot curve represents the ISR effect. As expected the fit of
real events (i.e. with ISR) is shifted towards higher value of the
mass. 
The two reconstructed curves also show a difference in width, which is
absent between the distributions in fig.~\ref{f1},
because of the extra smearing caused by the undetected ISR energy. 
The difference in width and mass is studied in table~\ref{tabisr}, also for
other values of $\sqrt{s}$. The table shows
shifts due to the ISR plus the neutrino reconstruction
($qqe\nu$ channel) around 300 MeV for the mass and 350 MeV for the width,
while the shifts due to the simple ISR (4q channel) are
about 50 MeV lower.

\begin{table}[hbt]\centering
\begin{tabular}{|c|c|c|c|c|c|c|}
\cline{2-7}
\multicolumn{1}{} {} & \multicolumn{2}{|c|}{172 GeV}& 
\multicolumn{2}{|c|}{184 GeV} &\multicolumn{2}{|c|}{200 GeV}\\
\cline{2-7} 
\multicolumn{1}{c|} {} & $\Delta m$ &  $\Delta \Gamma$   &
                         $\Delta m$ &  $\Delta \Gamma$   &
                         $\Delta m$ &  $\Delta \Gamma$ \\
\hline
$e\bar\nu u\bar d$ & 247 & 336 & 300 & 376 & 318 & 371  \\
\hline
4q       & 218 & 307 & 253 & 312 & 277 & 323   \\
\hline
\end {tabular}
\caption{Mass and width difference (in MeV) between distributions with and
  without ISR ($\Delta m=m_{ISR}-m_{noISR}$), for three values of 
\protect{$\sqrt{s}$} 
and for $qqqq$ and $qqe\nu$ channels.}
\label{tabisr}
\end {table}

\section{Conclusions}

We have investigated, from a theoretical point of view, the effects of
several approximations on the invariant mass distributions at LEP2
energies.
This study is particularly interesting in the light of the first
determinations of $m_W$ in LEP experiments using the direct 
reconstruction technique.

The distortion on the mass distributions have been studied both at
generator level and after a simulated reconstruction of the W mass
using a constrained fitting method.

Effects coming from CC20 interference, which produces a CC20-CC3 shift of about
40 MeV in m($e\nu$) and average distributions at parton level, 
are dominated by detector resolution, ISR and neutrino missing energy
effects after the reconstruction.
The interpretation of the single distortions is
much more complicated after the kinematic reconstruction because they 
become strictly connected with each other: ISR is linked to detector
inefficiencies and they have, in the case of semileptonic final state,   
strong effects on neutrino reconstruction.

Our studies, anyway, show that after reconstruction, detector and hadronization
effects mainly affect the width of the distributions.
The loss of information coming either from
undetected ISR or missing neutrino energy
has an effect on the mass shift of about  200 MeV and 30 MeV respectively at 
172 GeV; the first also  
 contributes to the broadening of the width.
 The distortions  increase
with $\sqrt{s}$.  This is due to purely kinematical
reasons for detector effects, and in general is a result of the
increasing missing energy in the event.
This introduces additional
complications  for a satisfactory determination of $m_W$, 
which can be
compensated only by a larger statistics.

\section*{Acknowledgments}
We would like to thank Niels Kjaer, author of PUFITC, for having
provided us with the program. We are indebted to Elena Accomando,
coauthor of WPHACT, and to Marco Bigi for useful discussions on the
arguments of this paper.


\vfill\eject

\vfill\eject

\vfill\eject

\vfill\eject

\end{document}